\documentclass[%
 reprint,
 superscriptaddress,
showpacs,preprintnumbers,
amsmath,
amssymb,
aps,
prd,
floatfix,
]{revtex4-1}

\usepackage{graphicx}% Include figure files
\usepackage{epstopdf}
\usepackage{dcolumn}% Align table columns on decimal point
\usepackage{bm}% bold math
\usepackage{color}
\usepackage{xspace}

\usepackage{hyperref}% add hypertext capabilities
\usepackage{natbib}
\usepackage{multirow}
\usepackage[displaymath, mathlines]{lineno}% Enable numbering of text and display math
\newcommand*\patchAmsMathEnvironmentForLineno[1]{%
  \expandafter\let\csname old#1\expandafter\endcsname\csname #1\endcsname
  \expandafter\let\csname oldend#1\expandafter\endcsname\csname end#1\endcsname
  \renewenvironment{#1}%
     {\linenomath\csname old#1\endcsname}%
     {\csname oldend#1\endcsname\endlinenomath}}% 
\newcommand*\patchBothAmsMathEnvironmentsForLineno[1]{%
  \patchAmsMathEnvironmentForLineno{#1}%
  \patchAmsMathEnvironmentForLineno{#1*}}%
\AtBeginDocument{%
\patchBothAmsMathEnvironmentsForLineno{equation}%
\patchBothAmsMathEnvironmentsForLineno{align}%
\patchBothAmsMathEnvironmentsForLineno{flalign}%
\patchBothAmsMathEnvironmentsForLineno{alignat}%
\patchBothAmsMathEnvironmentsForLineno{gather}%
\patchBothAmsMathEnvironmentsForLineno{multline}%
}

\input symbols
\linenumbersep{0.1cm}
\begin{document}
%\linenumbers

\title{First measurement of the muon neutrino charged current single pion production cross section on water with the T2K Near Detector}

%%%%%%%%%%%%%%%%%%%%%%%%%%%%%%%%%%%%%%%%%%%%%%%%%%%%%%%%%%%%%%
% T2K author list generated on Thu, 26 May 2016 01:38:51 +0900
% setting: extra = 0 revtex = 1 ptep = 0 simple = 0 xml = 0 yearrule = 1 shiftrule = 1
%         author list from archive (starting 2015/08/08 until now)
% Number of authors = 342
%%%%%%%%%%%%%%%%%%%%%%%%%%%%%%%%%%%%%%%%%%%%%%%%%%%%%%%%%%%%%%

\newcommand{\INSTEE}{\affiliation{University of Bern, Albert Einstein Center for Fundamental Physics, Laboratory for High Energy Physics (LHEP), Bern, Switzerland}}
\newcommand{\INSTFE}{\affiliation{Boston University, Department of Physics, Boston, Massachusetts, U.S.A.}}
\newcommand{\INSTD}{\affiliation{University of British Columbia, Department of Physics and Astronomy, Vancouver, British Columbia, Canada}}
\newcommand{\INSTGA}{\affiliation{University of California, Irvine, Department of Physics and Astronomy, Irvine, California, U.S.A.}}
\newcommand{\INSTI}{\affiliation{IRFU, CEA Saclay, Gif-sur-Yvette, France}}
\newcommand{\INSTGB}{\affiliation{University of Colorado at Boulder, Department of Physics, Boulder, Colorado, U.S.A.}}
\newcommand{\INSTFG}{\affiliation{Colorado State University, Department of Physics, Fort Collins, Colorado, U.S.A.}}
\newcommand{\INSTFH}{\affiliation{Duke University, Department of Physics, Durham, North Carolina, U.S.A.}}
\newcommand{\INSTBA}{\affiliation{Ecole Polytechnique, IN2P3-CNRS, Laboratoire Leprince-Ringuet, Palaiseau, France }}
\newcommand{\INSTEF}{\affiliation{ETH Zurich, Institute for Particle Physics, Zurich, Switzerland}}
\newcommand{\INSTEG}{\affiliation{University of Geneva, Section de Physique, DPNC, Geneva, Switzerland}}
\newcommand{\INSTDG}{\affiliation{H. Niewodniczanski Institute of Nuclear Physics PAN, Cracow, Poland}}
\newcommand{\INSTCB}{\affiliation{High Energy Accelerator Research Organization (KEK), Tsukuba, Ibaraki, Japan}}
\newcommand{\INSTED}{\affiliation{Institut de Fisica d'Altes Energies (IFAE), The Barcelona Institute of Science and Technology, Campus UAB, Bellaterra (Barcelona) Spain}}
\newcommand{\INSTEC}{\affiliation{IFIC (CSIC \& University of Valencia), Valencia, Spain}}
\newcommand{\INSTEI}{\affiliation{Imperial College London, Department of Physics, London, United Kingdom}}
\newcommand{\INSTGF}{\affiliation{INFN Sezione di Bari and Universit\`a e Politecnico di Bari, Dipartimento Interuniversitario di Fisica, Bari, Italy}}
\newcommand{\INSTBE}{\affiliation{INFN Sezione di Napoli and Universit\`a di Napoli, Dipartimento di Fisica, Napoli, Italy}}
\newcommand{\INSTBF}{\affiliation{INFN Sezione di Padova and Universit\`a di Padova, Dipartimento di Fisica, Padova, Italy}}
\newcommand{\INSTBD}{\affiliation{INFN Sezione di Roma and Universit\`a di Roma ``La Sapienza'', Roma, Italy}}
\newcommand{\INSTEB}{\affiliation{Institute for Nuclear Research of the Russian Academy of Sciences, Moscow, Russia}}
\newcommand{\INSTHA}{\affiliation{Kavli Institute for the Physics and Mathematics of the Universe (WPI), The University of Tokyo Institutes for Advanced Study, University of Tokyo, Kashiwa, Chiba, Japan}}
\newcommand{\INSTCC}{\affiliation{Kobe University, Kobe, Japan}}
\newcommand{\INSTCD}{\affiliation{Kyoto University, Department of Physics, Kyoto, Japan}}
\newcommand{\INSTEJ}{\affiliation{Lancaster University, Physics Department, Lancaster, United Kingdom}}
\newcommand{\INSTFC}{\affiliation{University of Liverpool, Department of Physics, Liverpool, United Kingdom}}
\newcommand{\INSTFI}{\affiliation{Louisiana State University, Department of Physics and Astronomy, Baton Rouge, Louisiana, U.S.A.}}
\newcommand{\INSTJ}{\affiliation{Universit\'e de Lyon, Universit\'e Claude Bernard Lyon 1, IPN Lyon (IN2P3), Villeurbanne, France}}
\newcommand{\INSTHB}{\affiliation{Michigan State University, Department of Physics and Astronomy,  East Lansing, Michigan, U.S.A.}}
\newcommand{\INSTCE}{\affiliation{Miyagi University of Education, Department of Physics, Sendai, Japan}}
\newcommand{\INSTDF}{\affiliation{National Centre for Nuclear Research, Warsaw, Poland}}
\newcommand{\INSTFJ}{\affiliation{State University of New York at Stony Brook, Department of Physics and Astronomy, Stony Brook, New York, U.S.A.}}
\newcommand{\INSTGJ}{\affiliation{Okayama University, Department of Physics, Okayama, Japan}}
\newcommand{\INSTCF}{\affiliation{Osaka City University, Department of Physics, Osaka, Japan}}
\newcommand{\INSTGG}{\affiliation{Oxford University, Department of Physics, Oxford, United Kingdom}}
\newcommand{\INSTBB}{\affiliation{UPMC, Universit\'e Paris Diderot, CNRS/IN2P3, Laboratoire de Physique Nucl\'eaire et de Hautes Energies (LPNHE), Paris, France}}
\newcommand{\INSTGC}{\affiliation{University of Pittsburgh, Department of Physics and Astronomy, Pittsburgh, Pennsylvania, U.S.A.}}
\newcommand{\INSTFA}{\affiliation{Queen Mary University of London, School of Physics and Astronomy, London, United Kingdom}}
\newcommand{\INSTE}{\affiliation{University of Regina, Department of Physics, Regina, Saskatchewan, Canada}}
\newcommand{\INSTGD}{\affiliation{University of Rochester, Department of Physics and Astronomy, Rochester, New York, U.S.A.}}
\newcommand{\INSTHC}{\affiliation{Royal Holloway University of London, Department of Physics, Egham, Surrey, United Kingdom}}
\newcommand{\INSTBC}{\affiliation{RWTH Aachen University, III. Physikalisches Institut, Aachen, Germany}}
\newcommand{\INSTFB}{\affiliation{University of Sheffield, Department of Physics and Astronomy, Sheffield, United Kingdom}}
\newcommand{\INSTDI}{\affiliation{University of Silesia, Institute of Physics, Katowice, Poland}}
\newcommand{\INSTEH}{\affiliation{STFC, Rutherford Appleton Laboratory, Harwell Oxford,  and  Daresbury Laboratory, Warrington, United Kingdom}}
\newcommand{\INSTCH}{\affiliation{University of Tokyo, Department of Physics, Tokyo, Japan}}
\newcommand{\INSTBJ}{\affiliation{University of Tokyo, Institute for Cosmic Ray Research, Kamioka Observatory, Kamioka, Japan}}
\newcommand{\INSTCG}{\affiliation{University of Tokyo, Institute for Cosmic Ray Research, Research Center for Cosmic Neutrinos, Kashiwa, Japan}}
\newcommand{\INSTGI}{\affiliation{Tokyo Metropolitan University, Department of Physics, Tokyo, Japan}}
\newcommand{\INSTF}{\affiliation{University of Toronto, Department of Physics, Toronto, Ontario, Canada}}
\newcommand{\INSTB}{\affiliation{TRIUMF, Vancouver, British Columbia, Canada}}
\newcommand{\INSTG}{\affiliation{University of Victoria, Department of Physics and Astronomy, Victoria, British Columbia, Canada}}
\newcommand{\INSTDJ}{\affiliation{University of Warsaw, Faculty of Physics, Warsaw, Poland}}
\newcommand{\INSTDH}{\affiliation{Warsaw University of Technology, Institute of Radioelectronics, Warsaw, Poland}}
\newcommand{\INSTFD}{\affiliation{University of Warwick, Department of Physics, Coventry, United Kingdom}}
\newcommand{\INSTGE}{\affiliation{University of Washington, Department of Physics, Seattle, Washington, U.S.A.}}
\newcommand{\INSTGH}{\affiliation{University of Winnipeg, Department of Physics, Winnipeg, Manitoba, Canada}}
\newcommand{\INSTEA}{\affiliation{Wroclaw University, Faculty of Physics and Astronomy, Wroclaw, Poland}}
\newcommand{\INSTH}{\affiliation{York University, Department of Physics and Astronomy, Toronto, Ontario, Canada}}

\INSTEE
\INSTFE
\INSTD
\INSTGA
\INSTI
\INSTGB
\INSTFG
\INSTFH
\INSTBA
\INSTEF
\INSTEG
\INSTDG
\INSTCB
\INSTED
\INSTEC
\INSTEI
\INSTGF
\INSTBE
\INSTBF
\INSTBD
\INSTEB
\INSTHA
\INSTCC
\INSTCD
\INSTEJ
\INSTFC
\INSTFI
\INSTJ
\INSTHB
\INSTCE
\INSTDF
\INSTFJ
\INSTGJ
\INSTCF
\INSTGG
\INSTBB
\INSTGC
\INSTFA
\INSTE
\INSTGD
\INSTHC
\INSTBC
\INSTFB
\INSTDI
\INSTEH
\INSTCH
\INSTBJ
\INSTCG
\INSTGI
\INSTF
\INSTB
\INSTG
\INSTDJ
\INSTDH
\INSTFD
\INSTGE
\INSTGH
\INSTEA
\INSTH

\author{K.\,Abe}\INSTBJ
\author{C.\,Andreopoulos}\INSTEH\INSTFC
\author{M.\,Antonova}\INSTEB
\author{S.\,Aoki}\INSTCC
\author{A.\,Ariga}\INSTEE
\author{S.\,Assylbekov}\INSTFG
\author{D.\,Autiero}\INSTJ
\author{S.\,Ban}\INSTCD
\author{M.\,Barbi}\INSTE
\author{G.J.\,Barker}\INSTFD
% ...

\collaboration{The T2K Collaboration}\noaffiliation

\date{\today}

\begin{abstract}
The T2K off-axis near detector, ND280, is used to make the first differential
cross section measurements of muon neutrino charged current single positive
pion production on a water target at energies ${\sim}0.8$~GeV.
The differential measurements are presented as a function of 
the muon and pion kinematics,
 in the restricted phase-space defined by
$\pipmom>200$\mevc, $\mumom>200$\mevc, $\pipcos>0.3$ and $\mucos>0.3$.
The total flux integrated \num charged current single positive pion production
cross section on water in the restricted phase-space is measured to be 
$\fluxav=4.25\pm0.48~(\mathrm{stat})\pm1.56~(\mathrm{syst})\times10^{-40}~\mathrm{cm}^{2}/\mathrm{nucleon}$. 
The total cross section is consistent with 
the NEUT prediction ($5.03\times10^{-40}~\mathrm{cm}^{2}/\mathrm{nucleon}$)
 and 2$\sigma$ lower than the GENIE prediction 
($7.68\times10^{-40}~\mathrm{cm}^{2}/\mathrm{nucleon}$). 
The differential cross sections are in good agreement with the
NEUT generator. 
The GENIE simulation reproduces well the shapes of the distributions, 
but over-estimates the overall cross section normalization.
\end{abstract}
\pacs{14.60.Pq, 14.60.Lm, 25.30.Pt, 29.40.Ka, 29.40.Mc}% PACS, the Physics and Astronomy
    % Need to check these.
\maketitle

\section{Introduction}
The T2K long baseline neutrino experiment~\cite{Abe:2011ks} has the primary goal to 
precisely measure neutrino oscillation parameters
 through measurements of \nue appearance and \num disappearance from a
 \num beam.
As neutrinos are charge-less and color-less, neutrino oscillation experiments rely on the detection
of charged particles coming from charged current (CC) and neutral
current (NC) interactions to infer neutrino properties, e.g. 
CC quasi elastic (QE) interactions allow the calculation of the neutrino
energy from the lepton kinematics.
The knowledge of \num and \nue interaction cross sections 
is then fundamental to infer neutrino properties correctly.
\num CC resonant interactions are part of the signal
and sometimes of the background of oscillation experiments, and a better
understanding of this channel could be beneficial not only to
T2K, but to the neutrino community in general, as there are
discrepancies between models and
experimental data.

Both the \mb~\cite{miniboone2011measurement} and 
\minerva~\cite{minerva2014charged} collaborations provided 
measurements of the CC single positive pion production (\ccpip) cross sections in mineral oil and 
plastic scintillator, respectively. 
The \ccpip cross section is described by the particles
 leaving the nucleus, i.e. one muon, one positive pion and any number of nucleons.
There are large discrepancies between the \mb and \minerva experiments, and
 the historic ANL~\cite{anl1982study} and BNL~\cite{bnl1986charged} 
 bubble chamber results,
 which could be due to nuclear effects that if not modeled correctly
 can modify the effective measured cross-section.
The \mb and \minerva results show also significant normalization
 and shape discrepancies between each other~\cite{minerva2014charged}, and currently no theoretical
 model can explain all the pion production data available.
Additional pion production data can help to constrain the pion
production models and give valuable information on the
nucleon-$\Delta$ axial form factor~\cite{PhysRevC.88.017601,PhysRevC.90.025501}.

We present the first \ccpip differential
cross section measurements on water.
A \ccpip measurement on water will have a strong impact on the 
T2K oscillation analysis, as current results suffer from large cross section
systematic uncertainties related to the differences in targets between
near and far detectors (carbon versus water)~\cite{LongOApaper}.
These data will also be beneficial to future atmospheric and long-baseline 
experiments, that plan to use a water target, such as 
the Hyper-Kamiokande experiment~\cite{abe2015physics}.

\section{T2K Experiment}
The T2K long baseline neutrino experiment uses the J-PARC facility in Tokai, Japan,  
to produce 30\,GeV protons, which
produce charged pions by colliding with a graphite target
and consequently result in a high purity \num beam.
The beam center axis is directed 2.5$^{\text{o}}$ off-axis towards 
Super-Kamiokande~\cite{Fukuda2003418} at 295\,km from J-PARC. 
Two near detectors are located at 280\,m from the target, the on-axis
near detector (INGRID~\cite{Otani2010368}) and the off-axis near 
detector (ND280).

{\it Neutrino Beam Flux}\textemdash 
The predicted neutrino beam flux~\cite{PhysRevD.87.012001} peaks at 0.6\,GeV
 and its fractional composition is
  92.6\% \num, 6.2\% \numb, 1.1\% \nue, 0.1\% \nueb. 
The proton interactions with the graphite target are simulated with
the FLUKA2008 package~\cite{battistoni2007fluka}, 
The propagation of secondary and tertiary pions and kaons and their 
decays to neutrinos is simulated with GEANT3~\cite{GEANT3}. 
The hadron interactions are modeled with GCALOR~\cite{GCALOR} and  tuned to
hadron production data from external experiments, such as
the CERN NA61/SHINE experiment~\cite{Abgrall:2011ae, Abgrall:2011ts, Abgrall:2015hmv}.

{\it Neutrino Interaction Model}\textemdash 
Based on the prediction of the neutrino flux, the NEUT~\cite{NEUT} (version 5.1.4.2)
event generator is used to simulate neutrino
interactions in ND280. 

For charged current quasi elastic (CCQE) and 
neutral current quasi elastic (NCQE) interactions, 
%For CCQE and NCQE interactions, 
NEUT uses the Llewellyn Smith model~\cite{smith1972ccqe}
integrated with the relativistic Fermi gas (RFG) 
model by Smith and Moniz to describe the nucleons
within the nucleus~\cite{smith1972rel}.
The outgoing nucleon is also required to have larger momentum than the
Fermi surface momentum (Pauli blocking), which is 217\mevc for carbon 
and 225\mevc oxygen.

NEUT uses the Rein-Sehgal model for resonant interactions~\cite{rein1981neutrino}, 
considering 18 resonances with masses below 2\gevc$^2$
and their interference terms.
In addition 20\% of the $\Delta$ resonances undergo pion-less $\Delta$
decay, in which the $\Delta$ is absorbed by the nuclear medium without
emitting any pions: $\Delta + N \to N' + N''$.
The NEUT pion production model is tuned using neutrino interaction
data from the MiniBooNE experiments~\cite{miniboone2011measurement,miniboone2011cc1pi0}, as explained in
Ref~\cite{LongOApaper}. 
In particular, the axial mass for resonant \ccpip interactions is set
to 1.41\gev, and the overall \ccpip normalization for energies less
than 2.5\gev is further increased by 15\% compared to 
predictions when the axial mass is set to 1.41\gev.

Coherent pion production is simulated for both NC and CC interactions
using the Rein-Sehgal model~\cite{rein1983coherent}, including the
PCAC (Partially Conserved Axial vector Current)
lepton mass correction for CC interactions~\cite{rein2007pcac}.

DIS (deep inelastic scattering) processes are simulated using GRV98 parton distribution 
functions~\cite{gluck1998dynamical} and corrections following the Bodek and Yang
model~\cite{bodek2003modeling} 
to improve the agreement with experiments in the low-$Q^2$ region.
To avoid double counting with the single pion resonant production, 
only multiple pion production processes are considered for the invariant mass of the recoiling hadron system
$W<2$\gevcc.
PYTHIA/JETSET~\cite{sjostrand1994pythia} is used for hadronisation
at energies above 2\gev, 
and an internal NEUT method is used at lower energies.

After the simulation of the initial neutrino-nucleon interaction, 
final state interactions are simulated with the cascade model~\cite{salcedo1988computer}.
Each particle is propagated inside the nucleus with steps
determined by the mean free path. 
The mean free path depends on the position inside the nucleus
and the momentum of the particle.
At each step, an interaction is generated according to the probability
calculated from each cross section such as charge exchange, absorption
or scattering.
If an interaction occurs, the resulting particles are used 
for stepping through the rest of the nucleus.
This process continues until all particles are either absorbed in the
nucleus or escape it.
Data from several pion scattering experiments are used to tune this model~\cite{LongOApaper}.

Additional information on the models used to simulate the neutrino interactions and
the hadron transport in the nuclear medium can be found
in references~\cite{NEUT,LongOApaper}.

The results in this paper are also compared to the GENIE
generator~\cite{Andreopoulos:2009rq}, 
as it provides a general framework valid over a large range of
experiments, targets and neutrino energy. 
GENIE uses essentially the same models as NEUT for the neutrino interactions
 simulation, but they differ in the implementation and value of some of the
parameters, such as $M_A^{RES}$ which is set to 1.12\gev~\cite{Kuzmin:2006dh}.
Single pion production (before final state interactions) comes from
resonant and coherent processes in NEUT, whereas GENIE also considers
DIS contributions to it. 
Although GENIE considers a lower value of $M_A^{RES}$, the predicted
single positive pion production cross-section is larger than in NEUT,
because DIS processes are allowed to contribute to this state.

{\it Near detector}\textemdash
ND280 is a complex of different sub-detectors enclosed in the 
refurbished UA1/NOMAD magnet. 
The origin of the ND280 coordinate system is at the center of the magnet
 and the 0.2\,T magnetic field is along the $+x$ direction. 
The $z$ axis is along the nominal neutrino beam axis, and $x$
and $y$ axes are horizontal and vertical, respectively.

\begin{figure}[!h]
  \includegraphics[trim=0cm 0cm 0cm 0cm, clip,
width=0.9\linewidth]{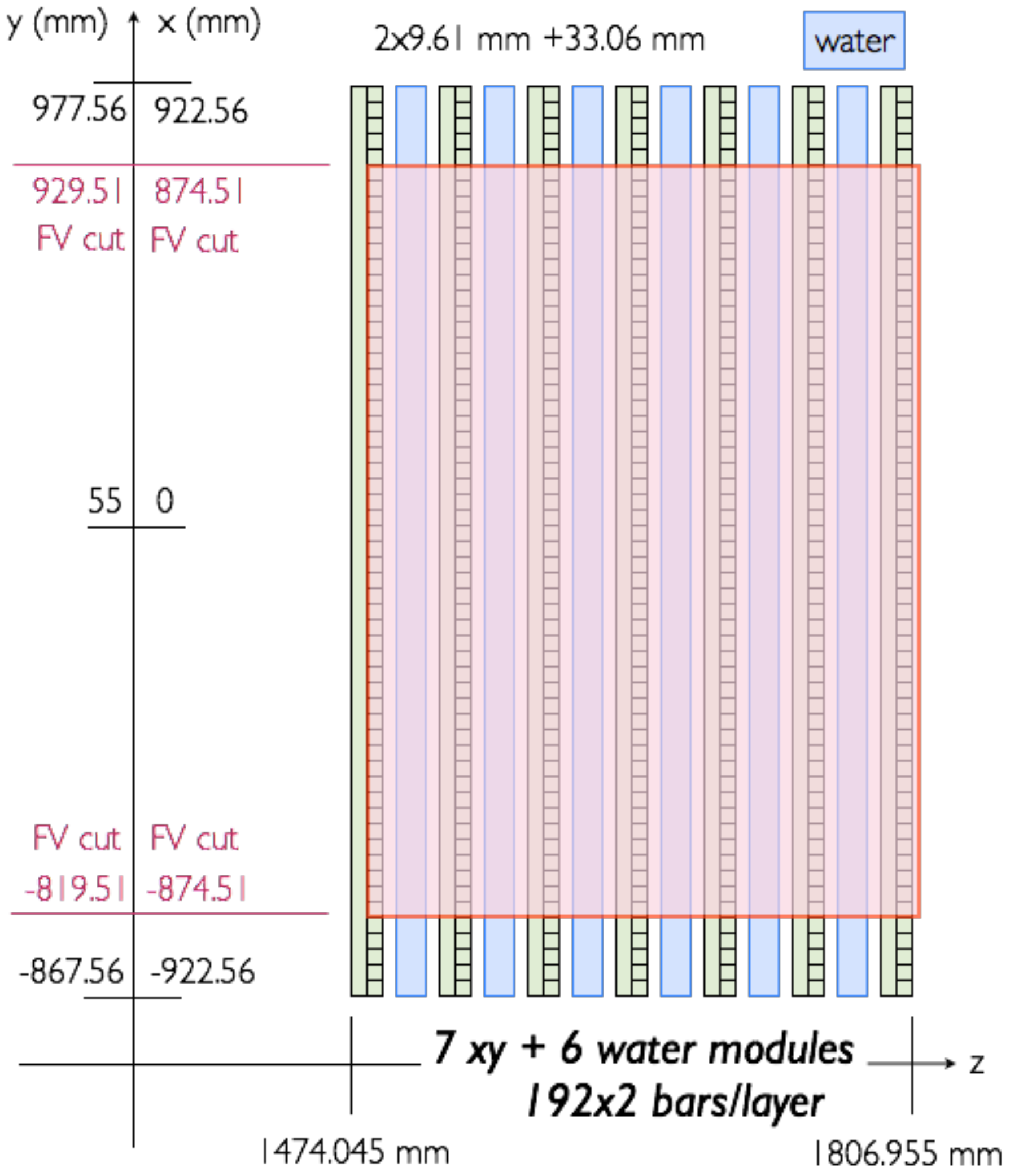}
\caption{\label{fig:fgd2structure} Schematic view of FGD2 and its fiducial volume (FV)
delimited by the red line. The first upstream scintillator layer is
not included in the FV.}
\end{figure}

The ND280 Tracker region contains two fine-grained detectors
 (FGDs~\cite{Amaudruz:2012pe})
 which are used as the neutrino interaction target, sandwiched between three gaseous 
time projection chambers
 (TPCs~\cite{Abgrall:2010hi}) which are used to track charged particles.
The most upstream FGD (FGD1) primarily consists of polystyrene scintillator 
bars with layers oriented alternately in the $x$ and $y$ 
directions allowing 3D tracking of charged particles.
The downstream FGD (FGD2) has a similar structure, but the polystyrene bars
are interleaved with water layers, creating a modular structure of
water layer + $x$ layer + $y$ layer + water layer, and so on (see Figure~\ref{fig:fgd2structure}).
The areal density of an $xy$ module and a water module are respectively $2146.3\pm14.4$\,mg/cm$^2$
and $2792.6\pm13.4$\,mg/cm$^2$.
This structure allows the measurement
of neutrino interactions on water.

The electromagnetic calorimeters (ECals~\cite{Allan:2013ofa}), made of layers
of lead and scintillator bars,
surround the Tracker region (Barrel-ECals) with one module downstream
of it (Ds-ECal). 
Upstream of the Tracker there is a $\pi^0$ detector (\POD~\cite{Assylbekov201248}), consisting
of scintillator, water and brass layers.

Magnet return yokes surround the entire detector to make the magnetic field uniform and
contain it inside the detector.
Plastic scintillators in the yoke form the side muon range detectors (SMRDs~\cite{Aoki:2012mf}).

The analysis here presented uses FGD2 as the active interaction target, where
in a signal event the neutrino interacts with a nucleus in the water layer, and the
charged lepton coming from a CC interaction is tracked in the 
downstream scintillator layers.
These results are based on data taken
from November 2010 to May 2013. 
The good quality data collected during this period corresponds to 
 $5.6 \times 10^{20}$ protons on target (\pot).

\begin{figure*}[t]
\includegraphics[width=0.45\linewidth]{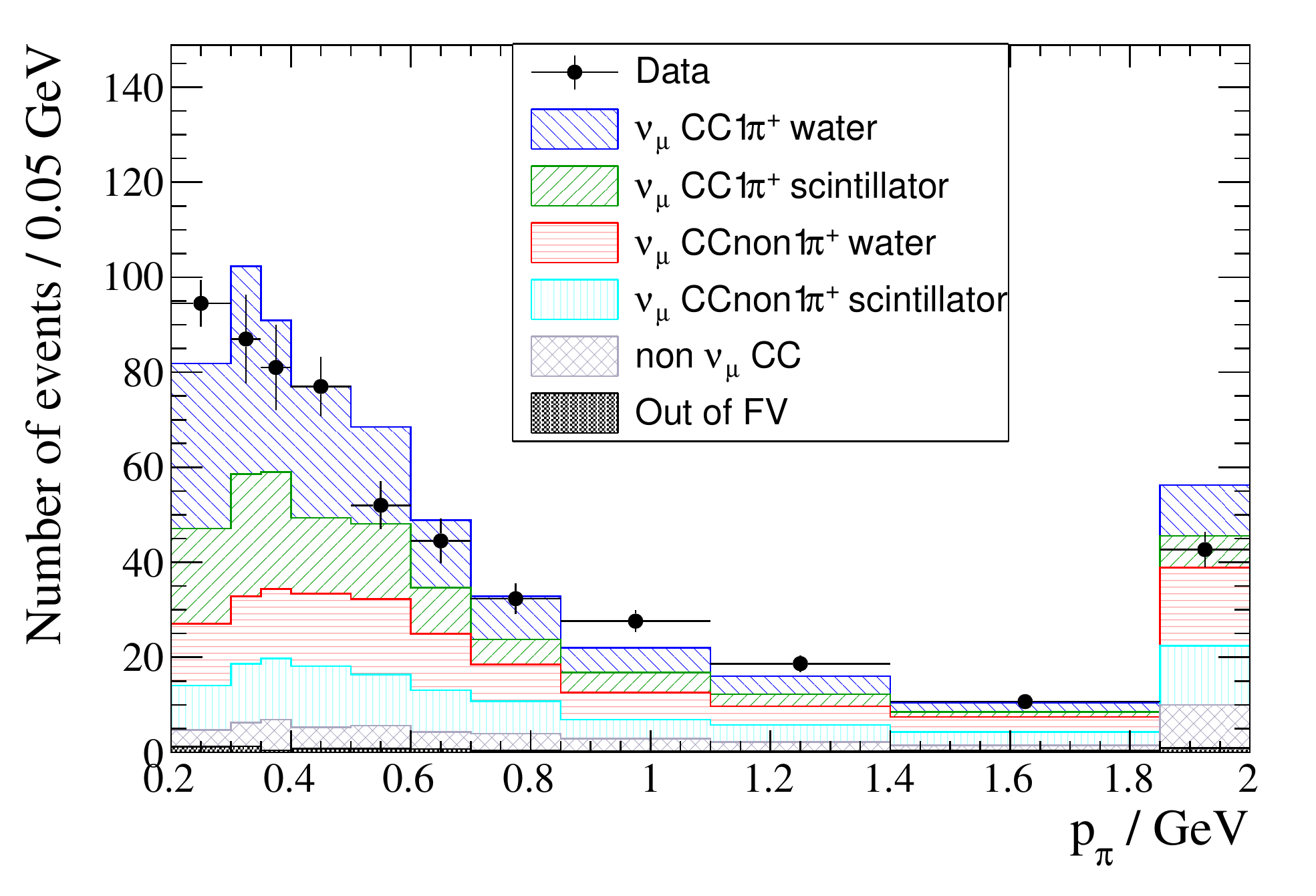} \,
\includegraphics[width=0.45\linewidth]{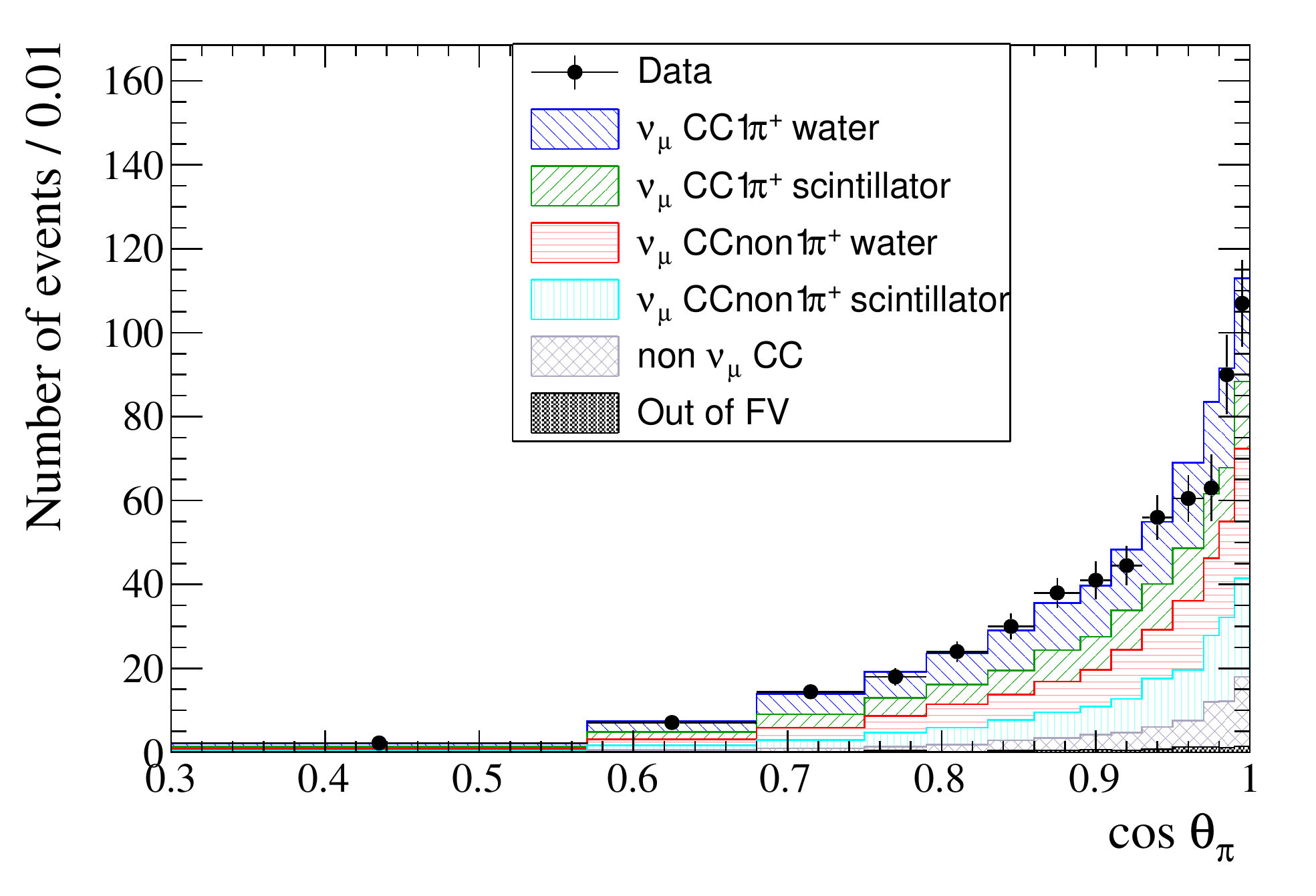} \\
\includegraphics[width=0.45\linewidth]{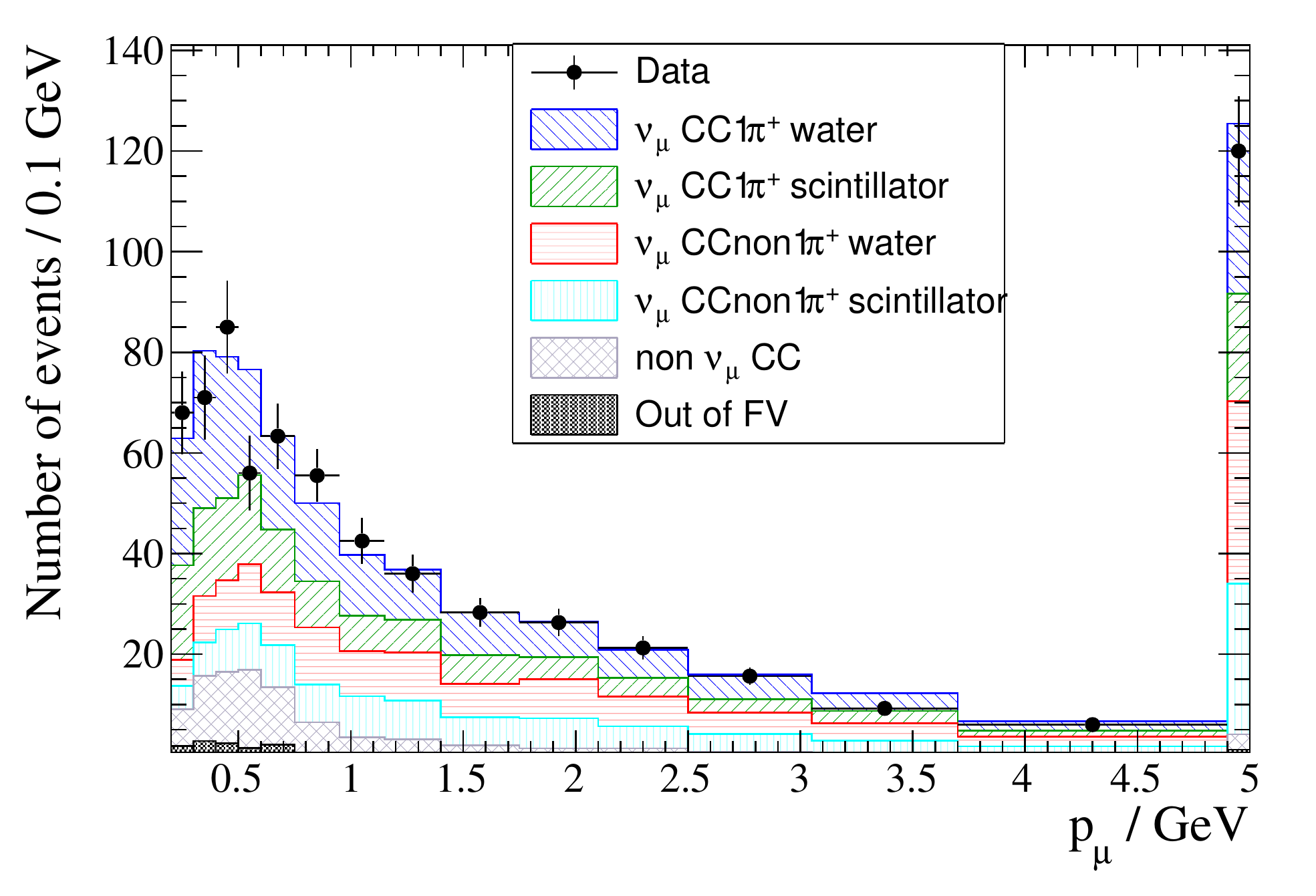} \,
\includegraphics[width=0.45\linewidth]{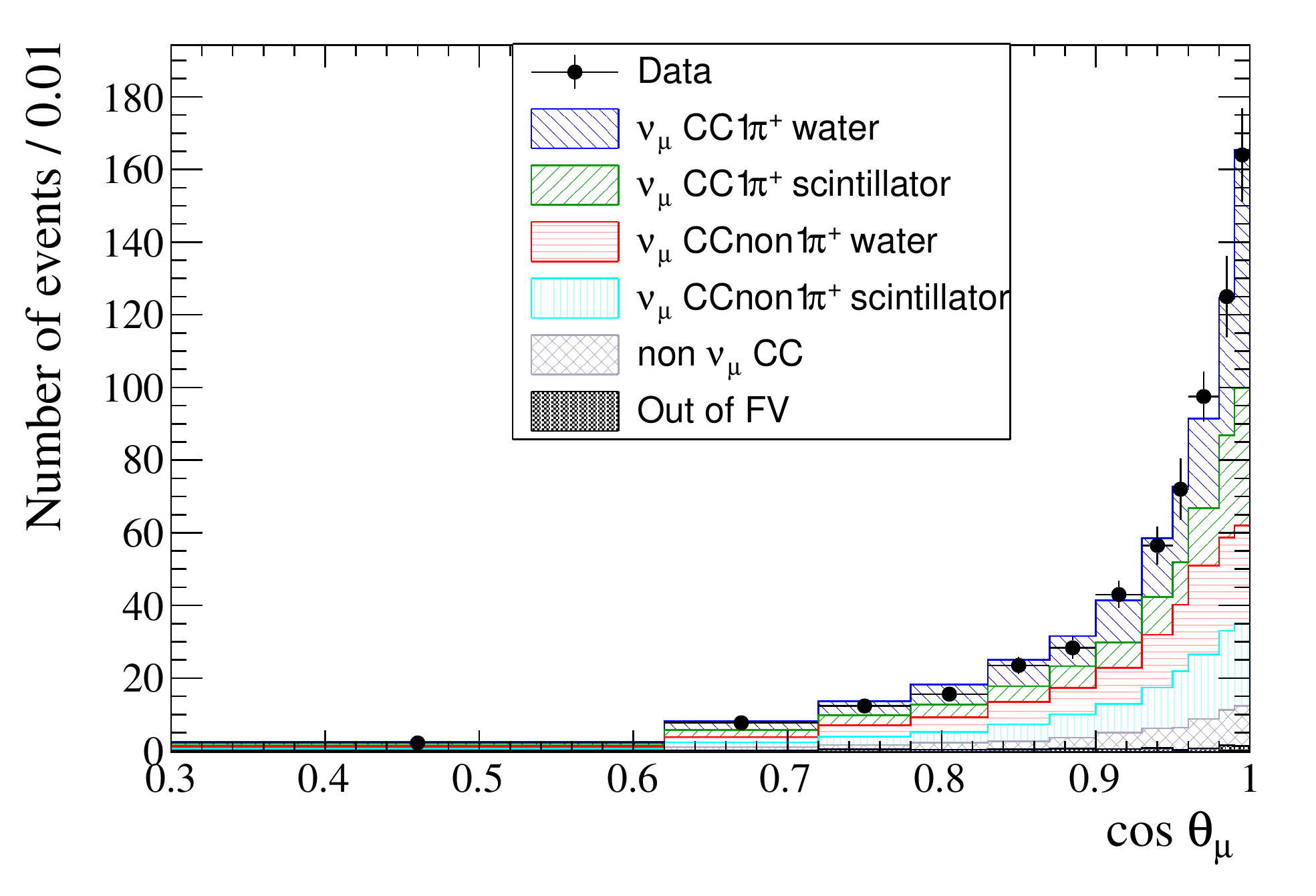} \\
\includegraphics[width=0.45\linewidth]{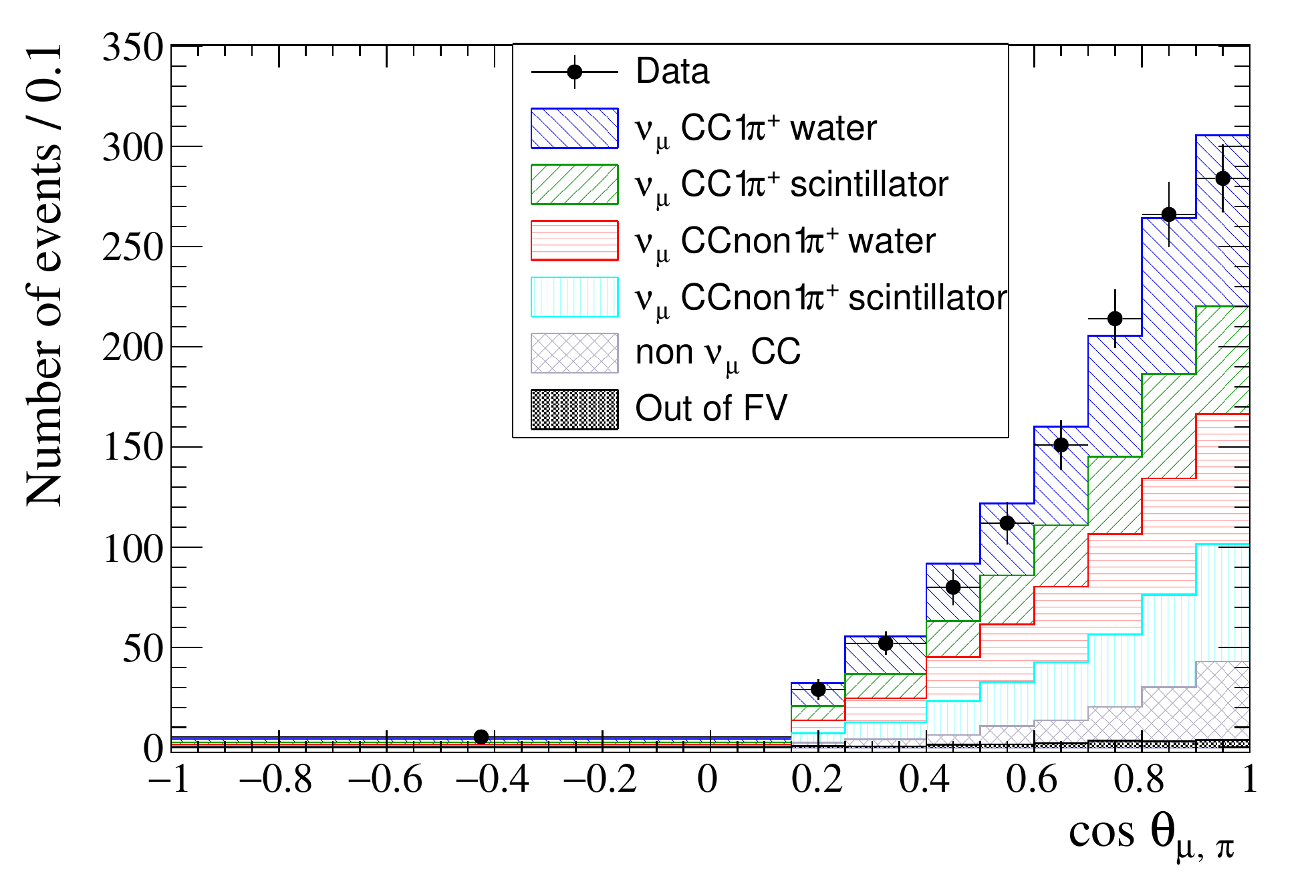} \,
\includegraphics[width=0.45\linewidth]{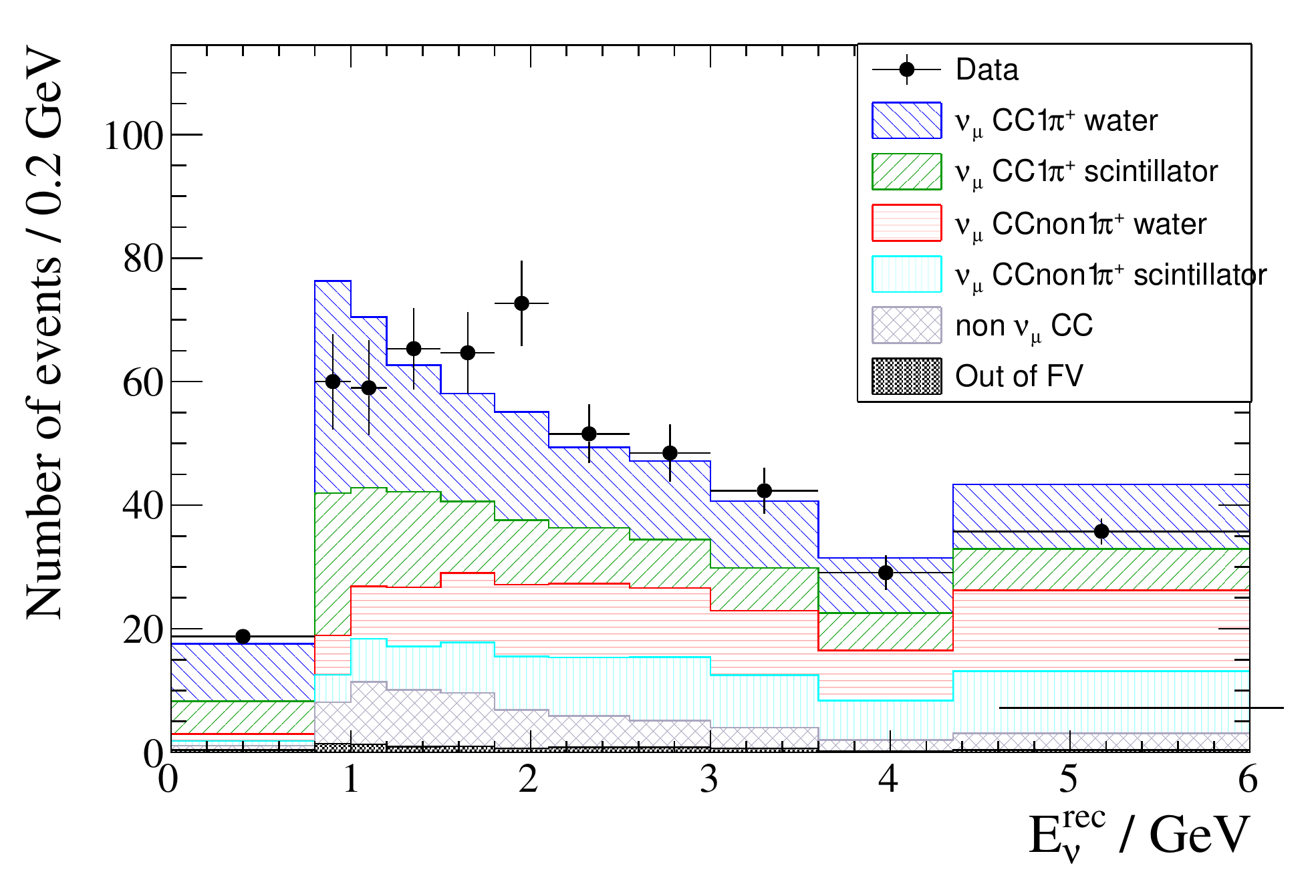} 
\caption{\label{fig:signal}Reconstructed pion kinematics (top), muon
  kinematics (middle), \mupipcos (bottom left) and neutrino energy
  (bottom right) distributions of the events in the \num \ccpip water-enhanced sample. The NEUT Monte Carlo prediction is separated into the 
\num \ccpip interactions in water, 
\num \ccpip interactions in scintillator, 
\num CCnon1\pip interactions in water, 
\num CCnon1\pip interactions in scintillator, 
non \num CC interactions, and interactions outside of the FGD2 FV.
The last bin in the \pipmom, \mumom and \erec distributions contains
all the over-flow events.
}
\end{figure*}

\begin{figure*}[t]
\includegraphics[width=0.45\linewidth]{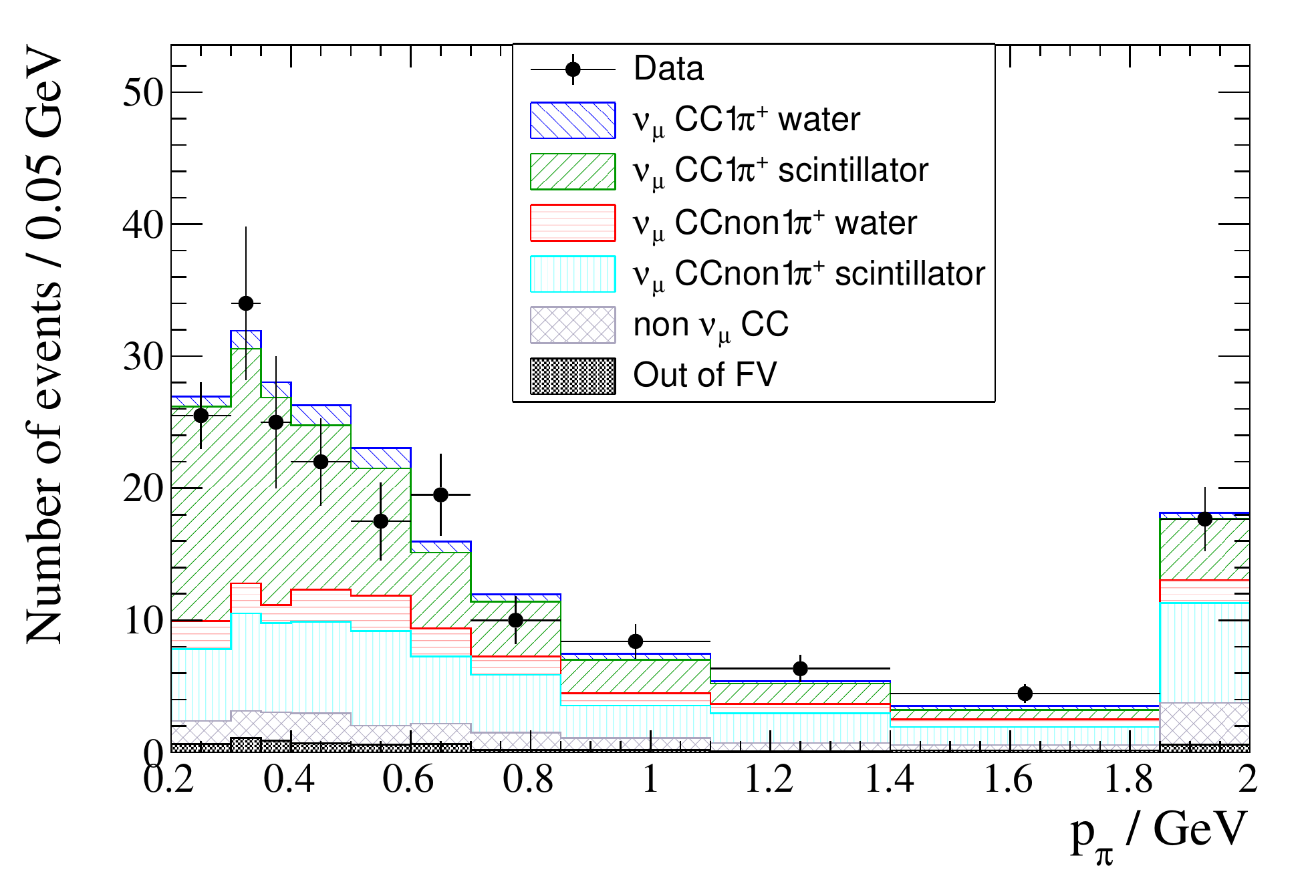} \,
\includegraphics[width=0.45\linewidth]{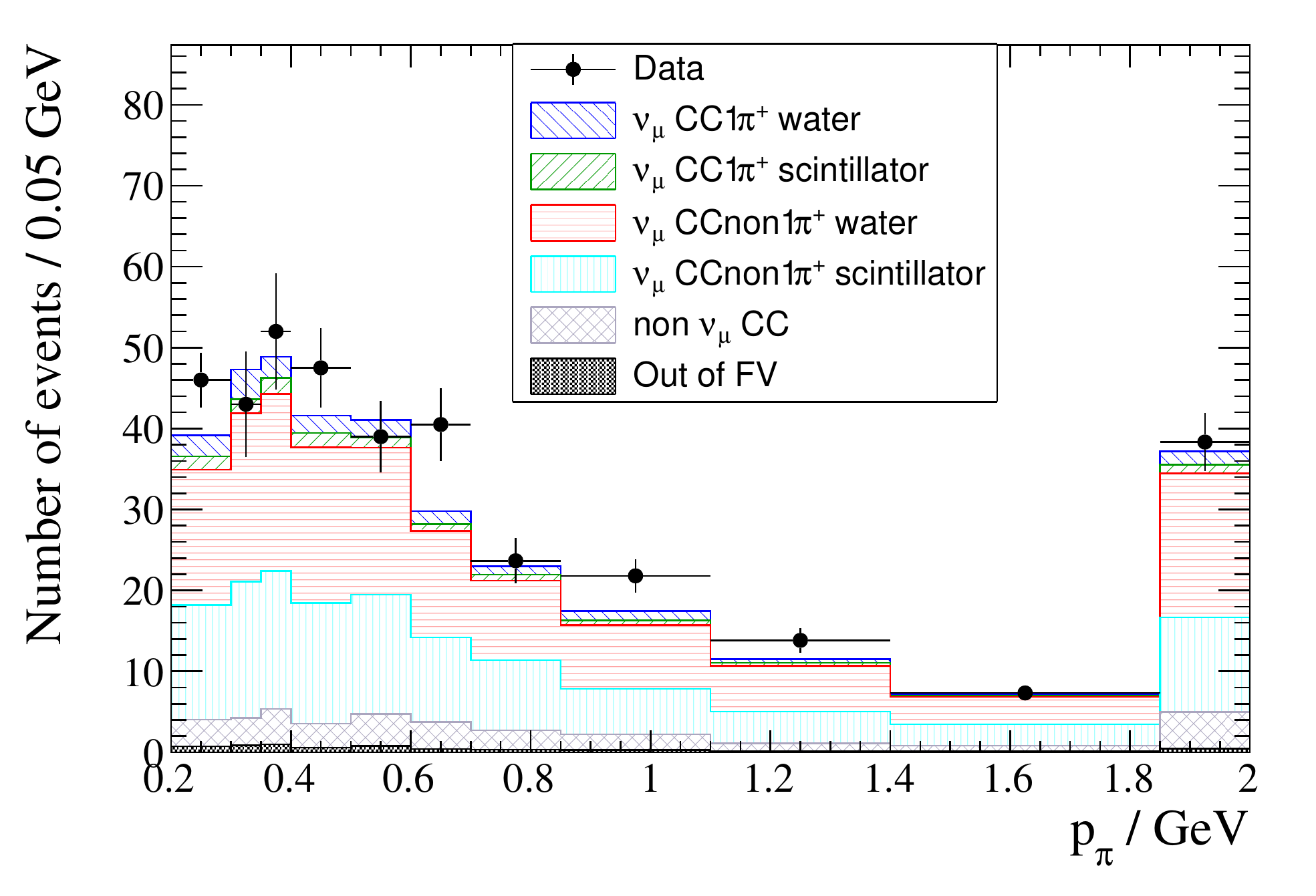} 
\caption{\label{fig:sideband}Reconstructed pion momentum
  distributions of the events in the 
 two external samples: \ccpip scintillator (left)
  and CC1$\pi^+$n$\pi$ water-enhanced (right).
The last bin contains all the over-flow events.
}
\end{figure*}

\section{Selection of \num \ccpip interactions in water}
Muon neutrino interactions are selected by using the highest momentum 
negative track starting in the fiducial volume (FV) of the FGD2.
The FGD2 FV begins 58\,mm inward from the lateral edges of the FGD2 and
7.5\,mm inward from the upstream FGD2 edge (as shown in Figure~\ref{fig:fgd2structure}).
These tracks are required to enter the TPC3 (located immediately downstream of FGD2)
and deposit energy
compatible with a muon-like track.
Additional tracks matched between the FGD and TPC associated with the same muon candidate vertex 
are tagged as either protons or positive, negative or neutral pions by looking at the trajectory and energy
deposit in the TPCs, and at electromagnetic showers in the ECals.
More details on the \num CC inclusive and multi-pion selections
can be found in References~\cite{NuMuCCInc,LongOApaper},
respectively, where the only differences are that in this analysis interactions
 in FGD2 are selected, rather than in FGD1, and the ECals are used 
to tag neutral pions.
\ccpip-like events are selected by requiring one muon, one positive 
pion, no other additional pions and any number of nucleons.

Because of the structure of the FGD2 (see Figure~\ref{fig:fgd2structure}),
 interaction vertices occurring in a water
module will be reconstructed in the $x$ layer downstream of it. 
A water-enhanced sample can be selected by requiring the vertex to be 
in the $x$ layer, whilst a scintillator sample can be selected by requiring
the reconstructed vertex to be in the $y$ layer.

The signal sample of this analysis is composed of 1402 selected \ccpip
 water-enhanced events in the full phase-space,
 with 30.9\% purity of true \num \ccpip 
interactions on water. 
To avoid relying on the simulation to describe regions of efficiency $<0.1$,
the analysis presented restricts the kinematic phase-space 
to the region defined by $\mumom > 200$\mevc, $\pipmom > 200$\mevc,
$\mucos >0.3$ and $\pipcos >0.3$.
With these restrictions in the phase-space of the signal definition, 
the signal efficiency goes from 13.3\% to 30.7\%.

The signal sample is selected with a purity of 39.9\,\% in the restricted phase-space.
Multi-pion interactions can be mis-identified as \ccpip interactions 
when one or more pions are absorbed by the detector or simply not reconstructed;
29.2\% of the signal sample is composed of this background.
\ccZpi interactions (3.9\%) come into the selection when the proton is
mis-identified as a \pip. 
The total background from \ccpip interactions occurring in the scintillator
amounts to 25.7\% of the signal sample,  
including interactions occurring in the $y$ layer whose vertex is reconstructed
in the $x$ layer.
Non-\num CC interactions (0.6\%) include both NC and CC interactions due to the
\nub and \nue components in the beam. 
They mainly come into the selection when a \pim from a NC 
interaction is mis-identified as the muon candidate.

\begin{table} 
  \caption{Percentage of true NEUT topologies (\ccpip and CCnon1\pip)
    in the restricted phase-space, and number of data events in different modules of the FGD2 for
  the signal (\ccpip water-enhanced) and two external samples (\ccpip scintillator 
  and CC1$\pi^+$n$\pi$ water-enhanced).}
  \begin{center} 
    \begin{tabular}{c|c|c|c}
                          &  \multicolumn{3}{c}{Selected samples} \\
                          & \ccpip & \ccpip & CC1$\pi^+$n$\pi$ \\
 True NEUT topology      & water  & scintillator & water \\
      \hline
           \ccpip water  & 39.9\,\% &    5.9\,\% &  7.7\,\% \\   
            \ccpip scint & 25.7\,\% &   54.6\,\% &  4.8\,\% \\   
      CC non1\pip water  & 18.5\,\% &    8.3\,\% & 49.0\,\% \\   
      CC non1\pip scint  & 14.6\,\% &   28.7\,\% & 36.5\,\% \\   
           non \num CC   &  0.6\,\% &    0.9\,\% & 1.2\,\% \\    
               Out of FV &  0.5\,\% &    1.7\,\% & 0.9\,\% \\    
      \hline
      Data in reduced phase-space &  1275   &    431    & 885 \\
      Data in full phase-space    &  1402   &    491    & 944 \\ 
    \end{tabular}
  \end{center}
  \label{tab:backgrounds}
\end{table}

These backgrounds are constrained with two external samples.
A sample of selected \num CC events with one \pip and at least one, 
but maximum 3, negative or neutral pions 
(CC1$\pi^+$n$\pi$ water-enhanced sample)
is used to constrain the non-\ccpip interactions, which include
the CC non1\pip, non \num CC and out of FV backgrounds.
A sample of selected \num \ccpip events in the $y$ layers of the FGD2
(CC1$\pi^+$ scintillator sample)
is used to constrain the background coming from the interactions in the 
scintillator.
Table~\ref{tab:backgrounds} shows the composition of the signal and external 
samples according to the NEUT generator.
Distributions of the pion kinematics (\pipmom and \pipcos), the muon
kinematics
(\mumom and \mucos), the cosine of the angle
between the muon and the pion (\mupipcos) 
and reconstructed neutrino energy (\erec) in the selected sample 
are shown in Figure~\ref{fig:signal}.
The reconstructed neutrino energy is found by applying 4-momentum conservation and 
assuming the target nucleon is at rest and the remaining final-state particle is a nucleon:
\begin{equation}
\erec = \dfrac{m_\mu^2 + m_\pi^2 - 2m_NE_f + 2 (p_\mu \cdot p_\pi) }
     {(2E_f - \mathbf{p_\mu} \mucos - \mathbf{p_\pi} \pipcos - m_N)}
\end{equation}
\noindent
where
$m_\mu$, $m_\pi$ and $m_N$  are the masses of the muon, the pion and the nucleon respectively;
$E_f = E_\mu + E_\pi$; 
$p_x$, $\mathbf{p_x}$ and $\theta_x$ are the 4-momentum, 3-momentum
and angle with the neutrino direction of the particle considered
($x=\mu, \pi^+$).
Distributions of the pion momentum in the external samples are found
in Figure~\ref{fig:sideband}.

{\it Systematic uncertainties}\textemdash
The TPC and FGD detector systematic uncertainties are the same as the ones
 described in References~\cite{NuMuCCInc,LongOApaper}.
The ECal particle identification systematic uncertainties are
evaluated with high purity samples of electrons and muons, as
explained in Reference~\cite{Abe:2014usb}.

The isolated ECal reconstruction systematic uncertainty is evaluated with a control sample
of both isolated and non-isolated ECal objects, due to the
difficulties of finding a control sample with just isolated ECal
objects. The efficiency is found to be 
$0.303\pm0.003$ in simulation and $0.315\pm0.009$ in data for the
Barrel-ECal, and
$0.826\pm0.002$ in simulation and $0.839\pm0.007$ for the Ds-ECal.
These efficiencies are used to correct the simulation efficiency for
tagging isolated-ECal objects only, which is 
$0.352$ for the Barrel-ECal and
$0.163$ for the Ds-ECal.

The FGD water modules mass uncertainty is 0.55\%.  
The FGD layer migration uncertainties have been evaluated 
in detail for this analysis.
These migrations are divided into forward  
(i.e. when the reconstructed vertex is a layer downstream of the
true vertex) and backward migrations (i.e. when the reconstructed vertex 
is a layer upstream of the true vertex). 
The forward migrations come from a hit reconstruction inefficiency.
Their overall uncertainty is estimated to be 3.3\% 
with a control sample of cosmic muons passing through both
FGDs. 
The backward migrations come from low energy backward going particles that 
are fitted with the muon track and move the vertex one or more layers
upstream. These latter migrations are estimated using the CC0$\pi$ and
CC multi-pion samples in FGD2: 
a normalization uncertainty of 30\% is assigned to them.

The flux uncertainties are evaluated with beam line and hadron production
measurements. 
The hadron production uncertainties dominate the neutrino
flux uncertainties, with a smaller contribution from the
neutrino beam direction and proton beam uncertainties.
The systematic uncertainty for the \num flux at ND280 
varies from 10\% and 15\% depending on the neutrino energy~\cite{PhysRevD.87.012001}.

The uncertainties related to the cross section model (final state interactions, 
CCQE model, pion production model and nuclear model) are constrained using
external data and comparisons between different existing models. 
A summary of these uncertainties can be found in Reference~\cite{LongOApaper}.

\section{Unfolding method}
The Bayesian unfolding technique by d'Agostini~\cite{DAgostini1995487} has been
successfully used by past T2K cross section measurements to extract
the cross sections (see References~\cite{NuMuCCInc,NuECCInc}).
The first estimate of the true distribution is found by applying
the unsmearing matrix $P(t_j|r_i)$ (found with Bayes' theorem)
 to the data distribution:
\begin{equation}
  \hat{N}_{t_j} = \dfrac{1}{\epsilon_j} \sum_i  P(t_j|r_i)(N_{r_i} - \sum_{k}^{\text{all backgrounds}}\alpha_k B_{r_i,k}) \; ,
\label{eq:unfolding}
\end{equation}
\noindent
 where  $t_j$ ($r_i$) indicates the true (reconstructed) bin for each observable,
 $N_{r_i}$ is the number of reconstructed events in bin $r_i$,
 $B_{r_i,k}$ is the number of predicted events in bin $r_i$ of background type $k$,
 $\alpha_k$ is a normalization constant derived from the external samples,
 and $\epsilon_j$ is the true efficiency in bin $t_j$.

Eq.~\ref{eq:unfolding} uses a background subtraction where
the coefficients $\alpha_k$ are 1 if that part of the background is
not constrained by any external sample, or otherwise calculated as:
\begin{equation}
  \alpha_k = \dfrac{C_{\text{data},k}}{C_{\text{MC},k}} \; ,
\label{eq:sideband}
\end{equation}
where $C_{\text{data},k}$ is the total number of events 
in external sample $k$ in data
and $C_{\text{MC},k}$ is the total number of events in external sample $k$ in MC. 
In this analysis the background is divided into 2 groups:
the CC1$\pi^+$ interactions in scintillator or in the
scintillator-like component of the water modules, that are constrained with the
CC1$\pi^+$ scintillator sample;
the non CC1$\pi^+$ background which is constrained with the CC1$\pi^+$n$\pi$ water-enhanced sample.

The FGD2 water modules are composed of
oxygen    (73.83\%),
carbon    (15.05\%),
hydrogen  (10.48\%),
silicon  (0.39\%), and
magnesium (0.25\%).
The carbon, silicon and magnesium come from the polycarbonate
structure that enclose the liquid water.
They compose the scintillator-like component of the water-modules and
can be subtracted out with the $x$-layer as they have similar composition.

The effect of the systematic uncertainties on the cross section 
measurements is evaluated by using pseudo-experiments. 
For each pseudo-experiment the signal and control samples are smeared according to the error
source considered, the normalization constants are re-evaluated and used to
re-normalize the signal prediction before evaluating the cross-section
for that throw.
The covariance matrix is then defined as:
\begin{equation}
  V_{ij}^s = \dfrac{1}{N}\sum_{s_n=1}^{N}(\sigma^{s_n}_i-\sigma^{nom}_i) (\sigma^{s_n}_j-\sigma^{nom}_j) \;,
\label{eq:covariance}
\end{equation}
where $\sigma^{s_n}_i$ is the differential cross section in bin $i$ evaluated with throw $n$ of the
uncertainty $s$, and 
$\sigma^{nom}_i$ is the nominal differential cross section in bin $i$.
Statistical and systematic uncertainties are evaluated by varying the 
contents of each bin according to Poisson and Gaussian statistics, respectively.

\begin{figure*}[t]
\includegraphics[trim=0cm 0.5cm 2.9cm 1.5cm, clip,
width=0.45\linewidth]{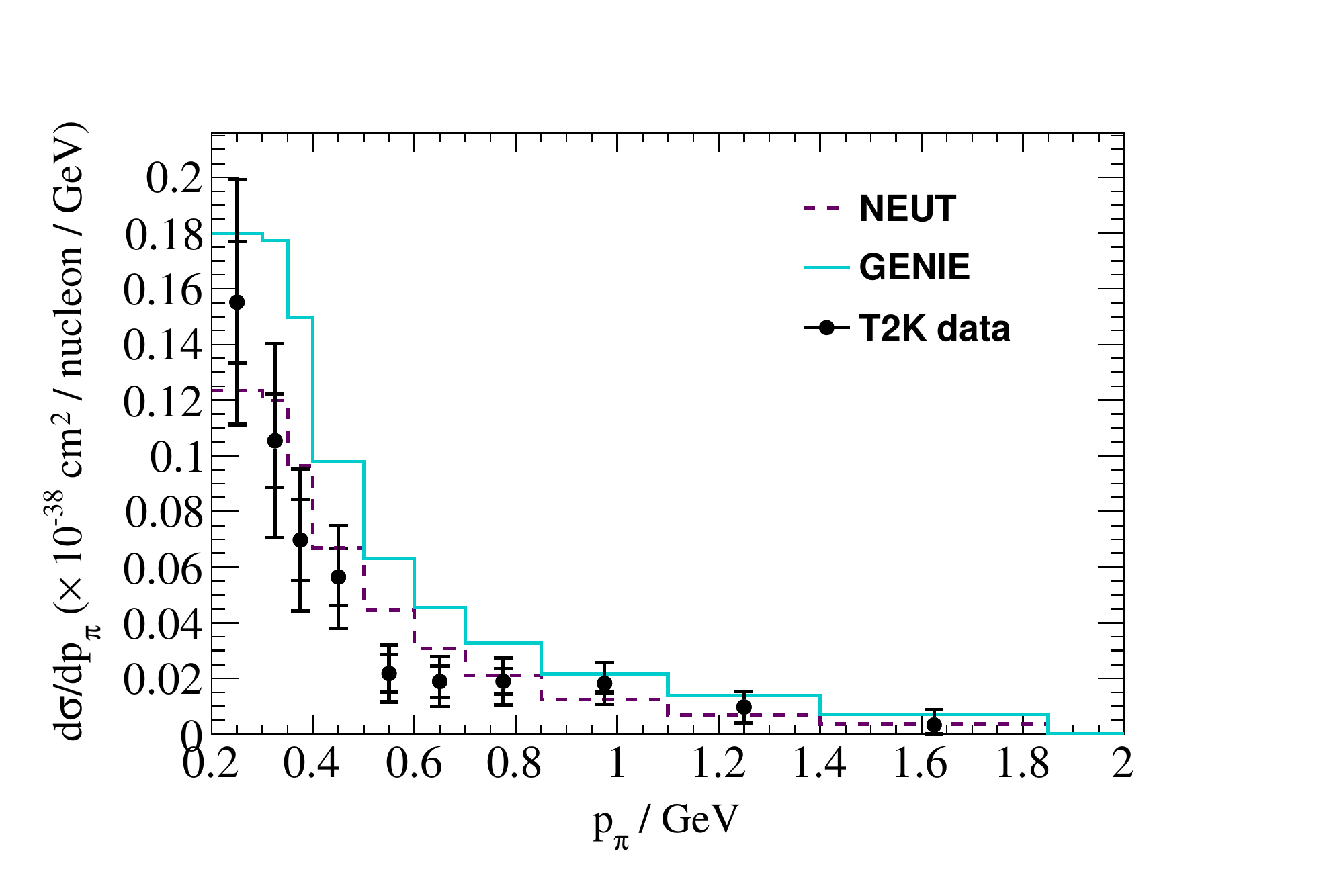} \,
\includegraphics[trim=0cm 0.5cm 2.9cm 1.5cm, clip,
width=0.45\linewidth]{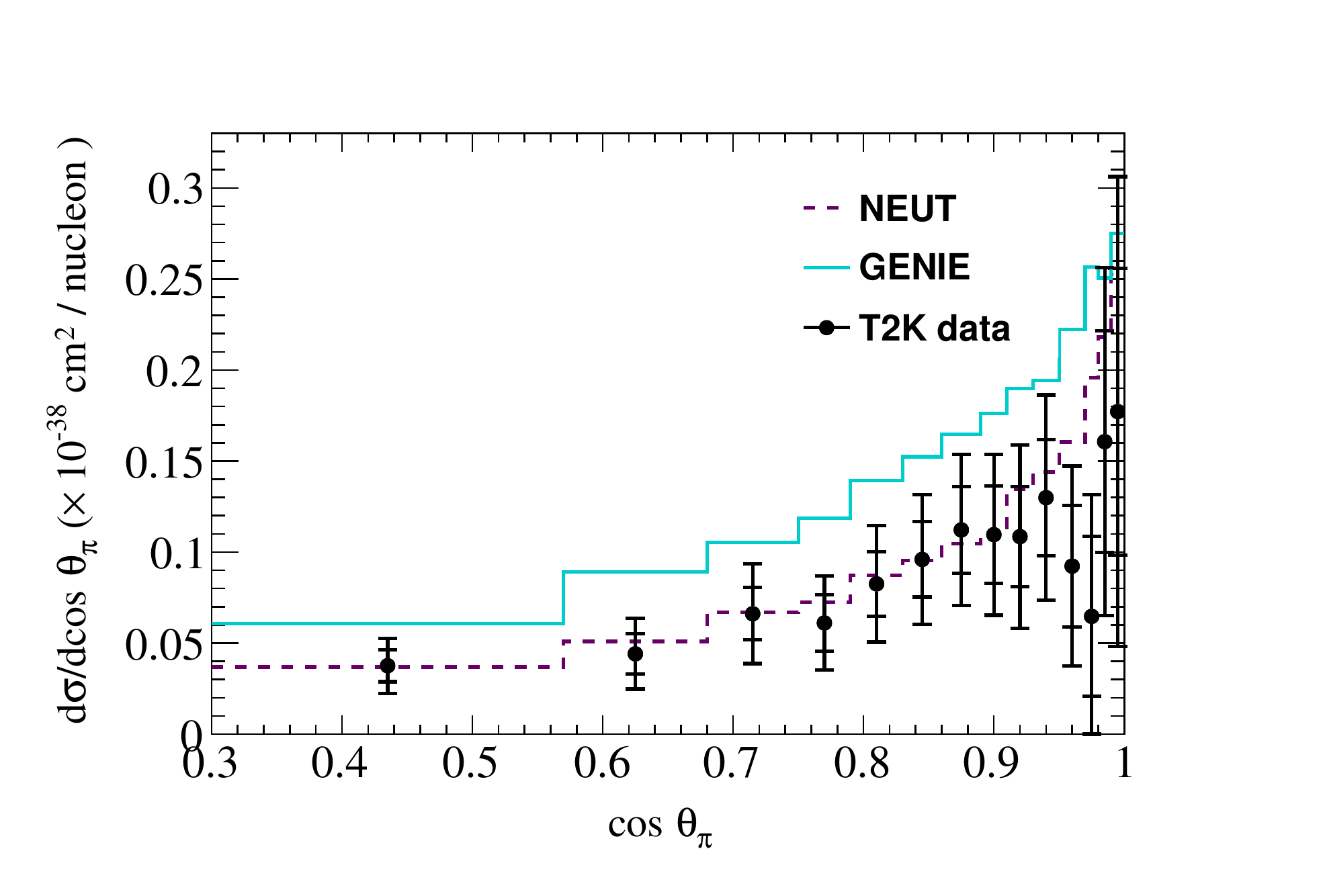} \\
\includegraphics[trim=0cm 0.5cm 2.9cm 1.5cm, clip,
width=0.45\linewidth]{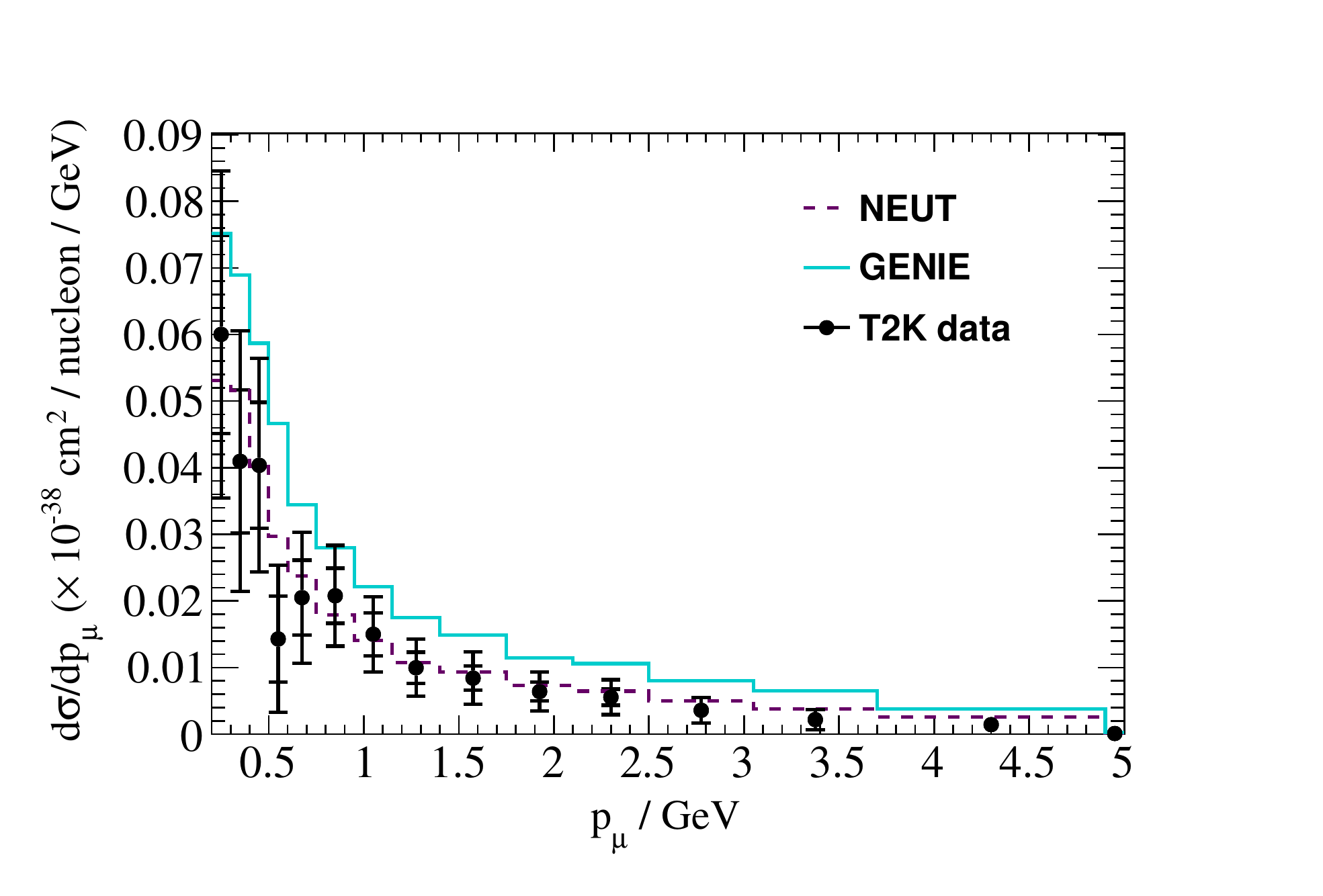} \,
\includegraphics[trim=0cm 0.5cm 2.9cm 1.5cm, clip,
width=0.45\linewidth]{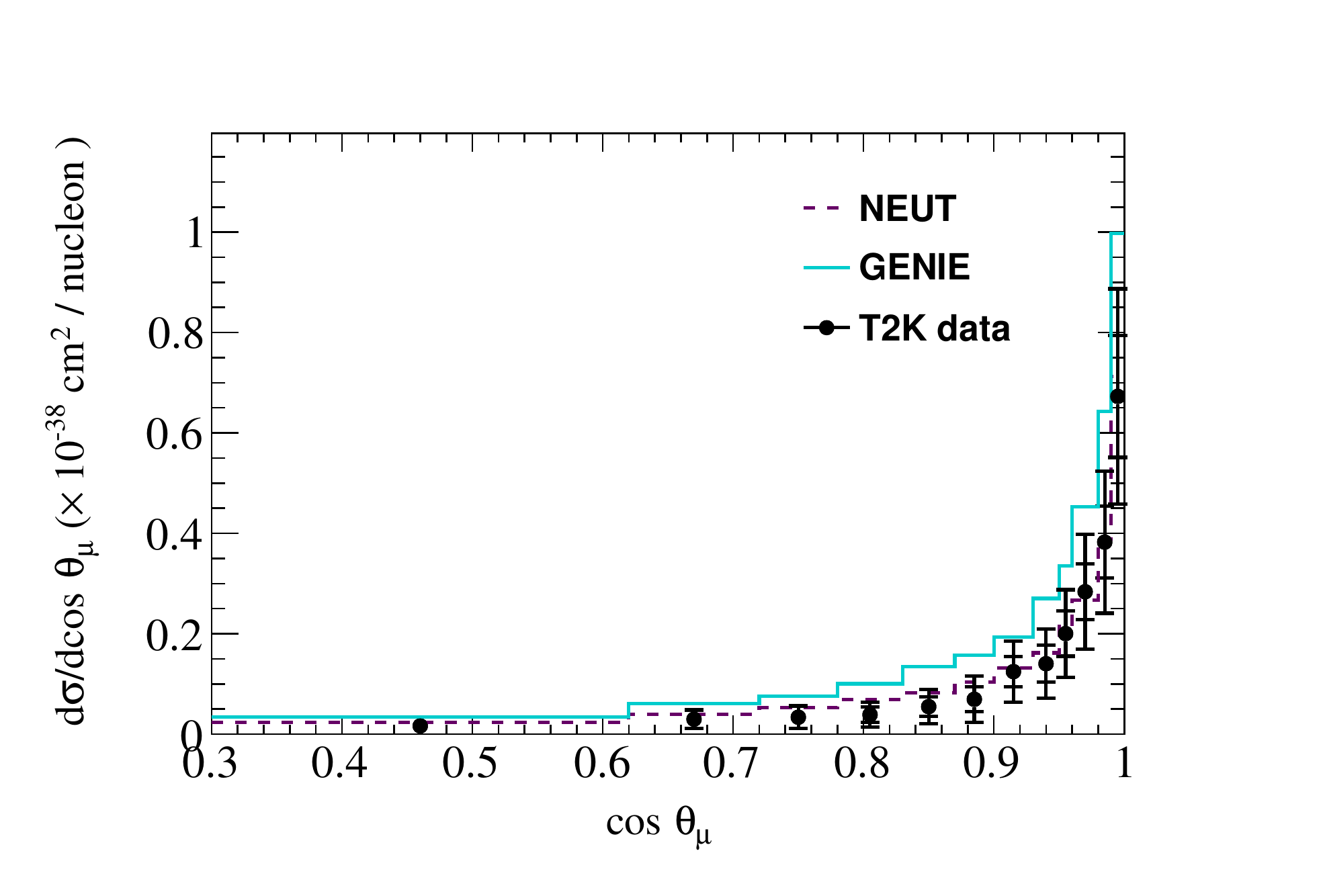} \\
\includegraphics[trim=0cm 0.5cm 2.9cm 1.5cm, clip,
width=0.45\linewidth]{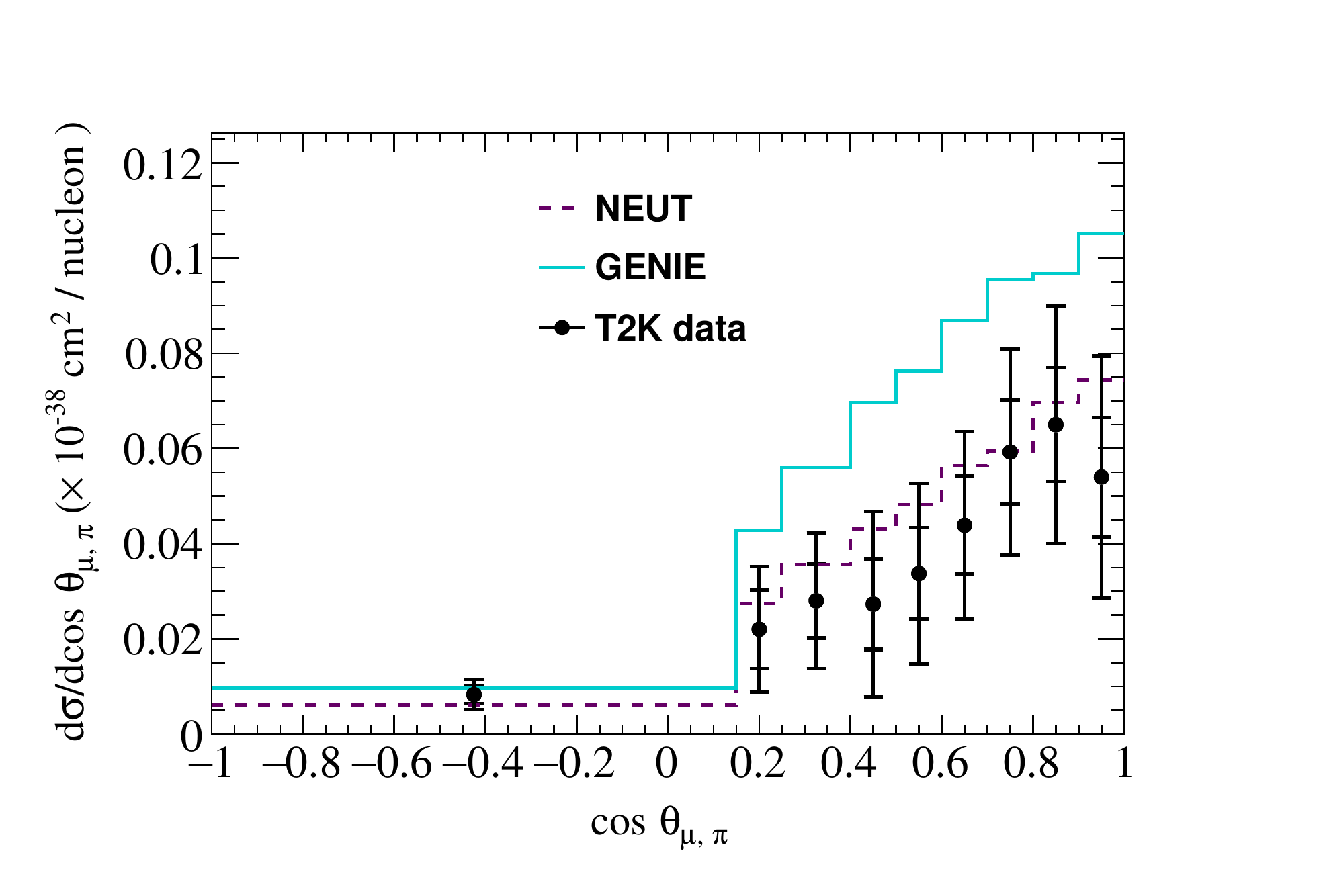} \,
\includegraphics[trim=0cm 0.5cm 2.9cm 1.5cm, clip,
width=0.45\linewidth]{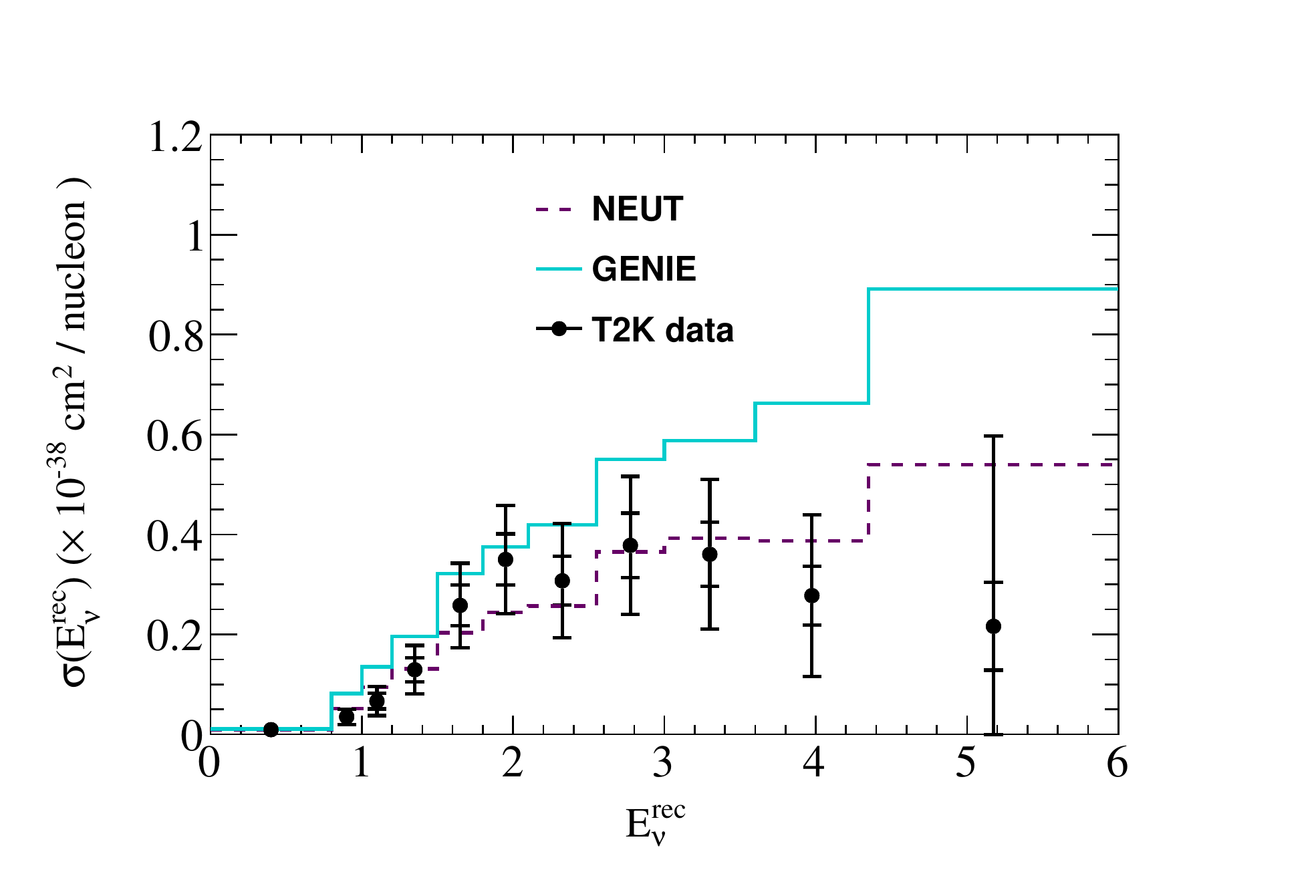} 
\caption{\label{fig:resultsdiff}Unfolded \num \ccpip differential
cross sections as a function of pion kinematics (top), muon kinematics
(center), \mupipcos (bottom left)
and \erec (bottom right) 
in the reduced phase-space of
$\pipmom>200$\mevc, $\mumom>200$\mevc, $\pipcos>0.3$ and $\mucos>0.3$.
For the \erec, the $\sigma(E)$ is presented as a model dependent result.
The inner (outer) error bars show the statistical (total) uncertainty on the data.
The dashed (solid) line shows the NEUT, version 5.1.4.2, (GENIE, version 2.6.4) prediction. }
\end{figure*}

\section{Cross-section results}
For a given variable $X$, the flux integrated differential cross section for bin $t_k$ is defined as:
 \begin{equation}
   \bigg\langle \dfrac{\partial \sigma}{\partial X} \bigg\rangle_{t_k} = \dfrac{N^{unfolded}_{t_k}}{T \Phi \Delta X_{t_k}} \, ,
   \label{eq:diffxsec}
 \end{equation}   
 \noindent
  where $N^{unfolded}_{t_k}$ is the estimated number of events in bin $t_k$ 
  (as given by Eq.~\ref{eq:unfolding}),
  $T$ is the number of target nucleons,  $\Phi$ is the \num flux
  per unit area and integrated over neutrino energy 
  (as detailed in Reference~\cite{PhysRevD.87.012001}),
  and $\Delta X_{t_k}$ is the width of bin $t_k$.
Even though single pion resonant production has a threshold at 480\,MeV, 
no cut is applied to the \num flux, as the \ccpip signal
definition includes processes with different thresholds as well.

The number of target nucleons is computed considering only the oxygen
and hydrogen in the FGD2 water modules, as the carbon, silicon and
magnesium components are removed by the Bayesian unfolding
with background subtraction. 
The total number of target nucleons is found to be:
\begin{equation*}
  T =  N_A \cdot V_{FV} \cdot \rho \displaystyle\sum_{\rm{a=O,H}} f_{a} \frac{A_{a}}{M_{a}} = 2.55 \cdot 10^{29} \; ,
\end{equation*}
  where $N_A = 6.022 \cdot 10^{23} mol^{-1}$ is the Avogadro number,
  $V_{FV}$ is the volume of the modules considered inside the FV,
  $\rho = \rho_{area}/\Delta z$ is the total density of the water modules of the FGD2
  ($\rho_{area} = 2798.7 \pm 5.4$ mg/cm$^2$ is the total areal density  
  and $\Delta z = 2.79$~cm is the width of each water module);
  $a$ runs over the elements present in the water modules; 
  $f_a$ is the mass fraction; $A_{a}$ represents the averaged number of nucleons per nucleus; and $M_a$ is the atomic mass.

The normalization constant found from the CC1$\pi^+$ control region is
$0.989 \pm 0.050$ indicating that the number of scintillator
interactions is compatible with the prediction from the simulation.
The normalization constant related to the non CC1$\pi^+$ background is $1.104 \pm 0.039$
indicating that the non CC1$\pi^+$ interactions are slightly more in data than simulation.
These two constants are used to re-normalize the background, and hence constrain the
systematic uncertainties.

Figure~\ref{fig:resultsdiff} shows the differential cross section as a function of 
pion kinematics (top), muon kinematics (center), \mupipcos (bottom left),
and \erec (bottom right). 
For the \erec the $\sigma(E)$ is presented as a model dependent result, as the \erec
is unfolded to the true neutrino energy as predicted by NEUT. 
The NEUT and GENIE (version 2.6.4~\cite{Andreopoulos:2009rq}) 
predictions are also shown in the plots.
The differential cross sections obtained are compatible with the NEUT prediction, 
but a small suppression is seen at $0.5<\pipmom<0.7$\,GeV and $\pipcos>0.95$.
This might be linked to the model for CC coherent interactions used in NEUT: NEUT greatly over-estimates the
amount of coherent interactions especially at low $E_{\pi^+}$~\cite{t2k2016coherent}.
The GENIE simulation reproduces well the shapes of the distributions, but over-estimates the overall 
cross section normalization.

\begin{figure}[t]
\includegraphics[trim=.8cm .8cm 1.2cm 1.2cm, clip, width=0.9\linewidth]{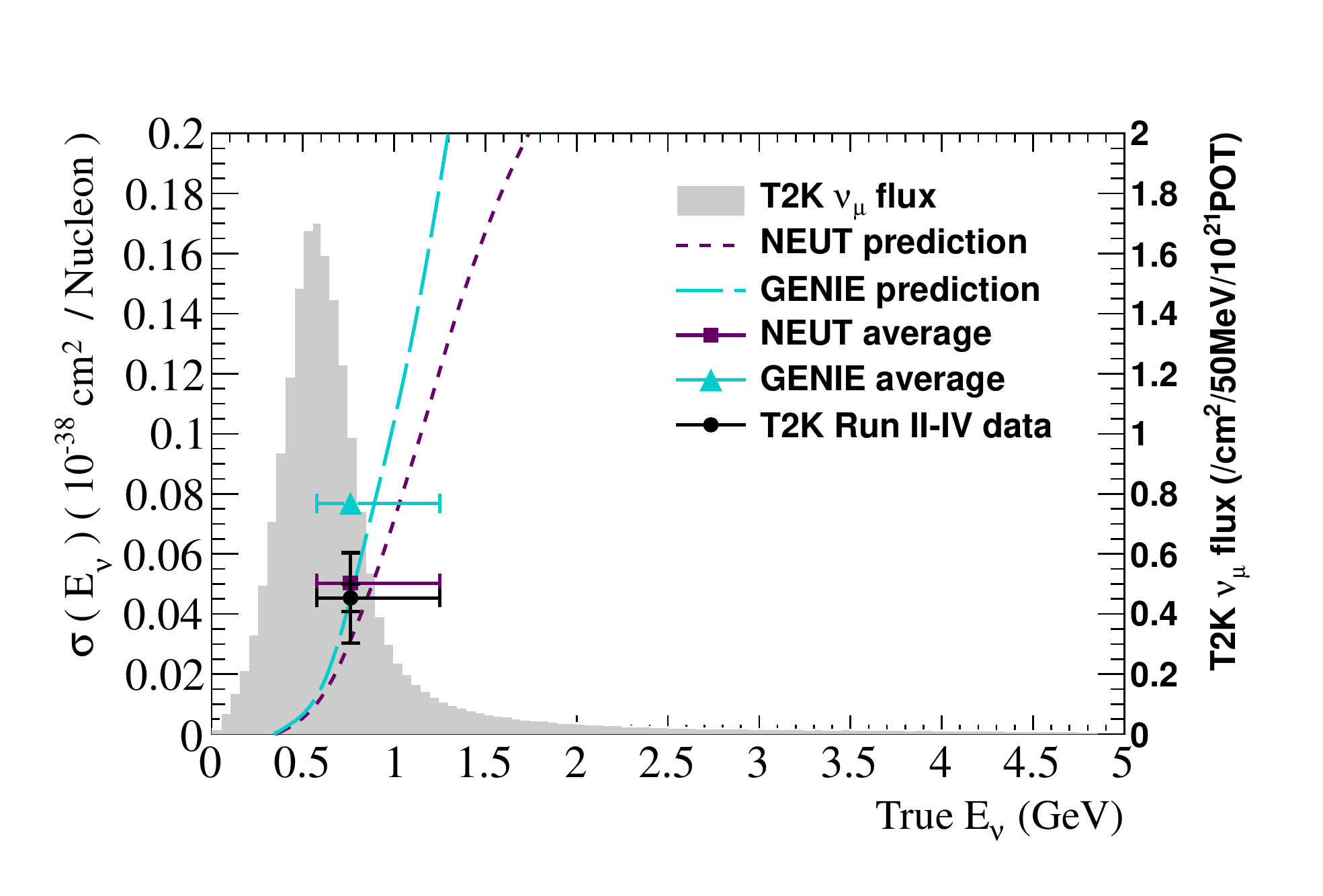}
\caption{\label{fig:resultstot} Total \num \ccpip cross section on water in the reduced phase-space of 
$\pipmom>200$\mevc, $\mumom>200$\mevc, $\pipcos>0.3$ and $\mucos>0.3$. 
The T2K data point is placed at the \num flux mean energy. 
The vertical error represents the total uncertainty, and the horizontal bar represents 68\% of the flux each
side of the mean. 
The T2K flux distribution is shown in gray. 
The NEUT (version 5.1.4.2) and GENIE (version 2.6.4) 
predictions are the total \num \ccpip 
predictions as a function of neutrino energy. The NEUT and
GENIE averages are the flux-averaged predictions. 
}
\end{figure}

The total flux integrated cross section is computed as:
\begin{equation}
  \langle \sigma \rangle_{\Phi} = \dfrac{N_{\text{total}}}{T \cdot \Phi} \;.
\label{eq:totalxsec}
\end{equation} 
The total flux integrated \num CC single positive pion production
cross section on water  in the restricted phase-space is measured to be 
$\fluxav=4.25\pm0.48~(\mathrm{stat})\pm1.56~(\mathrm{syst})\times10^{-40}~\mathrm{cm}^{2}/\mathrm{nucleon}$. 
This result is compatible with the NEUT prediction of $5.03\times10^{-40}~\mathrm{cm}^{2}/\mathrm{nucleon}$, 
and about 2\,$\sigma$ away from the GENIE prediction $7.68\times10^{-40}~\mathrm{cm}^{2}/\mathrm{nucleon}$.
The dominant systematic uncertainties on this result are those related
to the cross section model (23.9\%)
and flux parameters (25.5\%), because of the low purity of the selected
signal sample.
Without the selected control samples both these uncertainties would be
as high as 40\%. 
Nonetheless the low statistics and purity of the selected control
samples makes it difficult to further reduce these uncertainties.
%The flux uncertainties are larger than the initial uncertainty size
%because of the low purity of the selected signal sample.
Final state interactions and detector systematic uncertainties contribute with 
5.3\% and 10.8\%, respectively.
The data and MC statistical errors are estimated as 10.7\% and 3.3\%, respectively.
Figure~\ref{fig:resultstot} shows the total \num \ccpip cross section on water in the reduced phase-space of 
$\pipmom>200$\mevc, $\mumom>200$\mevc, $\pipcos>0.3$ and $\mucos>0.3$, with the T2K \num flux and the NEUT and GENIE predictions.

%In the future the flux uncertainties should be reduced by using more data 
%from the NA61 experiment in the flux prediction.
%The theory cross-section systematic uncertainties could be reduced by unfolding
%the signal and control regions simultaneously. This technique will allow to 
%control the signal and other backgrounds leaking in the control regions.
%As more data will be available, the data statistical uncertainties will also be
%reduced in the future.
%The detector systematic uncertainties are dominated by the pion secondary interactions, 
%TPC PID and FGD forward migration systematics. 
%

Future analyses will consider the use of the FGD2 and FGD1 samples simultaneously, 
eliminating the necessity to divide the FGD2 sample in the $x$ and $y$
layers and allowing the simultaneous evaluation of the cross-sections on
scintillator and water. This technique will
considerably reduce both the flux and theory cross-section
uncertainties that currently limit this measurement.

The data related to this measurement can be found together with the cross section
results obtained when unfolding the muon kinematics and neutrino energy distributions 
in Reference~\cite{Abe:datarelease}.

{\it Conclusion}\textemdash
The T2K off-axis near detector ND280 is used to extract the first \num \ccpip
differential cross sections on water as a function of the pion kinematics and muon-pion angle.
These results will be beneficial to the T2K experiment and the 
neutrino community in general, as a better understanding of neutrino induced
pion production on water at energy below 2\gev would result in a higher sensitivity
to the measurement of oscillation parameters.
The cross section is evaluated in the restricted phase-space defined by
$\mumom>200$\mevc, $\pipmom>200$\mevc, $\mucos>0.3$ and $\pipcos>0.3$.
The results are in good agreement with the NEUT generator % and we report 
%a suppression at $0.5<\pipmom<0.7$\,GeV and $\pipcos>0.95$.
and a general suppression is seen compared to the GENIE generator.
The total \num \ccpip cross section on water is found to be 
$\fluxav=4.25\pm0.48~(\mathrm{stat})\pm1.56~(\mathrm{syst})\times10^{-40}~\mathrm{cm}^{2}/\mathrm{nucleon}$,
 which is in good agreement with the NEUT
prediction and is 2$\sigma$ lower than the GENIE prediction.

{\it Acknowledgements}\textemdash
We thank the J-PARC staff for superb accelerator performance and the 
CERN NA61 Collaboration for providing valuable particle production data.
We acknowledge the support of MEXT, Japan; 
NSERC (Grant No. SAPPJ-2014-00031), NRC and CFI, Canada;
CEA and CNRS/IN2P3, France;
DFG, Germany; 
INFN, Italy;
National Science Centre (NCN), Poland;
RSF, RFBR and MES, Russia; 
MINECO and ERDF funds, Spain;
SNSF and SERI, Switzerland;
STFC, UK; and 
DOE, USA.
We also thank CERN for the UA1/NOMAD magnet, 
DESY for the HERA-B magnet mover system, 
NII for SINET4, 
the WestGrid and SciNet consortia in Compute Canada, 
and GridPP in the United Kingdom.
In addition, participation of individual researchers and institutions has been further 
supported by funds from ERC (FP7), H2020 Grant No. RISE-GA644294-JENNIFER, EU; 
JSPS, Japan; 
Royal Society, UK; 
and the DOE Early Career program, USA.

\bibliographystyle{apsrev4-1}
\bibliography{NuMuCC1piPRD}

\end{document}